\documentclass[aps,prd,reprint,twocolumn,nofootinbib]{revtex4-1}
\pdfoutput=1 

\usepackage{graphicx}
\usepackage{bm}
\usepackage{amsmath}
\usepackage{amssymb}
\usepackage{amsfonts}
\usepackage{amsthm}
\usepackage{mathtools}
\usepackage{hyperref}
\usepackage{braket}
\usepackage[normalem]{ulem}
\usepackage{wrapfig}
\usepackage{tikz}
\usepackage{dsfont}
\usepackage{comment}

\hypersetup{colorlinks=true,urlcolor=[rgb]{0,0,0.5},citecolor=[rgb]{0.5,0,0},linkcolor=[rgb]{0,0,0.4}}

\def\beq{\begin{equation}}
\def\eeq{\end{equation}}

\begin{document}
\title{Smeared end-of-the-world branes}

\author{William Harvey}
\email{wharvey@uvic.ca}
\author{Kristan Jensen}
\email{kristanj@uvic.ca}
\author{Takahiro Uzu}
\email{takahirouzu@uvic.ca}

\affiliation{\it Department of Physics and Astronomy, University of Victoria, Victoria, BC V8W 3P6, Canada}

\begin{abstract}
We uncover new non-supersymmetric boundary conditions in 10- and 11-dimensional supergravity whereby spacetime ends on a smeared distribution of D- and M-branes respectively. For example, we find a solution of type IIB supergravity where the AdS$_5\times\mathbb{S}^5$ vacuum ends on an $SO(6)$-invariant distribution of D3-branes. These distributions give a stringy completion of simple models of tensionful end-of-the-world branes considered previously in the literature. However we find that our solutions are all unstable to the fragmentation of the end-of-the-world brane into its constituents.
\end{abstract}

\maketitle

%--------------------------------------
\emph{Introduction.}~How can spacetime end? This is an old question whose answer is relevant in big bang cosmologies as well as in string theory and holography. In this work we study how spacetime can end on a dynamical object, an ``end-of-the-world (ETW) brane,'' in the context of string theory realizations of the AdS/CFT correspondence. In particular, we consider the holographic duals of boundary conformal field theories (BCFTs) where the ETW brane is dual to the boundary of the BCFT.

In addition to being ingredients in the holographic dictionary, ETW branes are also useful tools for studying aspects of the black hole information problem. They appear in the ``West Coast'' description of islands in~\cite{Penington:2019kki}; some non-generic microstates are dual to black hole spacetimes with an ETW brane behind the horizon~\cite{Cooper:2018cmb}; and a cousin of ETW branes, domain walls between two different regions of spacetime, are an important ingredient in the islands description of the Page curve of evaporating black holes~\cite{Almheiri:2019psf,Penington:2019npb,Almheiri:2019qdq,Penington:2019kki,Geng:2020qvw,Chen:2020uac}.

There are several known string theory realizations of supersymmetric ETW branes. For example, within AdS/CFT, there are supersymmetric BCFTs given by parallel D3 branes ending on a network of five-branes~\cite{Gomis:2006cu,DHoker:2007zhm,DHoker:2007hhe}, and M2 branes ending on M5 branes~\cite{Bachas:2013vza}, holographically dual to intricate supergravity backgrounds. The former is a fibration which is locally AdS$_4\times\mathbb{S}^2\times\mathbb{S}^2\times\Sigma$ with $\Sigma$ a strip together with a three-form flux background and axiodilaton which vary over $\Sigma$, with a non-trivial five-dimensional interpretation~\cite{Huertas:2023syg}.

Because these proper string theory realizations of ETW branes are rather complicated, most recent efforts in the literature focus on toy models of ETW branes, going back to~\cite{Takayanagi:2011zk} (building on~\cite{Karch:2000ct} and related works), in which spacetime ends on a tensionful brane, perhaps with matter living on it, in such a way as to respect the Israel junction condition. This simple model has been embedded into string theory in realizations where the ETW brane is an orientifold or a pair of orientifolds~\cite{Fujita:2011fp}, but generally it is an ad hoc model. 

In this Letter we attempt to embed this simple model into the lamppost examples of AdS/CFT, namely the AdS$_5\times\mathbb{S}^5$ vacuum of type IIB supergravity (SUGRA), and the AdS$_4\times\mathbb{S}^7$ and AdS$_7\times\mathbb{S}^4$ vacua of M-theory. The main players in our supergravity realizations of ETW branes are the Ramond-Ramond (RR) fluxes that support Freund-Rubin compactifications like the AdS$_5\times\mathbb{S}^5$ vacuum of type IIB SUGRA, and the compact or ``internal'' space in the compactification, in that case the $\mathbb{S}^5$. We study backgrounds with rotational symmetry on the internal space, and where the remaining non-compact space has a boundary, an ETW brane. That boundary must be charged to be consistent with RR flux ending on it, i.e. it carries $N$ units of 3-brane charge, and the rotational symmetry then implies that the brane charge sourcing that flux must be homogeneously distributed over the internal space. This distribution of brane charge leads to a modified version of the Israel junction condition, with a non-compact part that can be interpreted as the junction condition for the ETW brane, and a component in the internal directions that amounts to force balance on it. In the context of the $\mathbb{S}^5$ compactification of type IIB SUGRA, the ETW brane is a smeared distribution of 3-branes, while in the $\mathbb{S}^7$ and $\mathbb{S}^4$ compactifications of 11d supergravity we have smeared distributions of 2- and 5-branes respectively. 

We uncover solutions consistent with these boundary conditions in all three of these cases, focusing on the $\mathbb{S}^5$ compactification of type IIB SUGRA in the body of this Letter and leaving the 11d SUGRA examples to the Appendix. All of the solutions we find rely upon a nontrivial warpfactor for the internal space, although numerically the solutions are nearly AdS$_p\times \mathbb{S}^q$ up until very close to the ETW brane. Our solutions are all new, although a domain wall similar to our type IIB SUGRA solution was briefly considered in~\cite{Kraus:1999it}. Moreover our solutions are weakly curved and therefore within the regime of validity of bulk effective field theory. 

In a companion paper~\cite{constraints} we show how to meaningfully compare top-down and bottom-up models of ETW branes. That analysis builds on~\cite{Reeves:2021sab}, using the fact that holographic BCFTs are in a sense special: they have approximate singularities in correlation functions corresponding to, in the bulk, null trajectories that bounce off the ETW brane. Those trajectories are characterized by a quantity we call $\phi_b$, which allows us to determine if a top-down ETW brane can be modeled by a bottom-up one, and if so, to deduce its tension.

We find that our smeared ETW brane can be modeled by such a bottom-up model in AdS$_5$ with a positive-tension ETW brane. However we also find that our top-down model is unstable to fragmentation, whereby the individual branes that make up the smeared ETW brane are unstable to fluctuations in their transverse directions. 
%present an argument (building on~\cite{Reeves:2021sab}) for a criterion for whether a holographic ETW brane is sensible as a low-energy effective theory, but does not admit a completion into a theory of quantum gravity. The solutions we uncover fail that criterion, as does the ad hoc model~\cite{Takayanagi:2011zk} and many generalizations thereof. Consistent with that argument, we find that our solutions are unstable to fragmentation, whereby the individual branes that make up the smeared ETW brane are unstable to fluctuations in their transverse directions. 

A lesson from this work is that the internal space cannot be ignored in top-down constructions of ETW branes. 

The remainder of this Letter is organized as follows. We begin with a brief review of type IIB SUGRA and its $SO(6)$ invariant compactification on a $\mathbb{S}^5$. We then allow for spacetime to end on a smeared distribution of 3-branes and uncover the appropriate generalization of the Israel junction condition. We then present an asymptotically AdS$_5\times\mathbb{S}^5$ solution consistent with the ETW brane boundary condition, and demonstrate that it is unstable. In our discussion we present a candidate brane interpretation and boundary condition dual to the ETW brane and relegate our 11d examples to the Appendix.

%--------------------------------------
\emph{Type IIB SUGRA compactified on $\mathbb{S}^5$}.~In the body of this Letter we study Type IIB SUGRA. It has an AdS$_5\times\mathbb{S}^5$ solution supported by $N=\frac{L^4}{4\pi\ell_s^4}$ units of self-dual five-form flux through a $\mathbb{S}^5$,
\begin{align}
\begin{split}
    ds_{10}^2 &= L^2(ds^2_{\rm AdS_5} + d\Omega_5^2) \,, \\
    F_5 &= 4L^4 \left( \text{vol}_{\rm AdS_5} + \text{vol}_{\mathbb{S}^5}\right)\,,
\end{split}
\end{align}
where $ds_{10}^2$ is the 10d Einstein-frame line element, $\ell_s$ is the string length, and $F_5$ is the Ramond-Ramond five-form flux. Implicitly the 10d axiodilaton $\tau$ is constant and all other fields vanish. Here $ds^2_{\rm AdS_5}$ is the line element on a unit radius AdS$_5$, $\text{vol}_{\rm AdS_5}$ its volume form, $d\Omega_5^2$ is the line element on a round unit radius $\mathbb{S}^5$ and $\text{vol}_{\mathbb{S}^5}$ its volume form.

Consider the $SO(6)$ invariant compactification of type IIB SUGRA on a $\mathbb{S}^5$ with $N$ units of D3-brane flux through it. The Einstein-frame metric and five-form flux for such a background are
\begin{align}
    ds_{10}^2 &= e^{-\frac{10}{3}\phi}ds^2_{5} + e^{2\phi}d\Omega_5^2\,, 
    \\
    \nonumber
    F_{\mu\nu\rho\sigma\tau} &= 4L^4 e^{-\frac{40}{3}\phi} \varepsilon_{\mu\nu\rho\sigma\tau}\,,\qquad F_{\alpha\beta\gamma\delta\epsilon} = 4L^4 \varepsilon_{\alpha\beta\gamma\delta\epsilon}\,.
\end{align}
Here $ds_5^2$ refers to a 5d metric $g_{\mu\nu}(x^{\rho})$, $\mu,\nu,..$ to 5d indices and $\alpha,\beta,..$ to angles on the $\mathbb{S}^5$. The warpfactor in front of the $\mathbb{S}^m$ is parameterized by $\phi(x^{\mu})$. We consistently set the axiodilaton to a constant and the three-form fluxes to vanish. In terms of 5d quantities the 10d equations for such a geometry can be recast as 5d Einstein's equations with an appropriate stress tensor and a Klein-Gordon like equation for the warpfactor $\phi$,
\begin{align}
\begin{split}
\label{E:EinsteinAndScalar}
	R_{\mu\nu} - \frac{R}{2}g_{\mu\nu}& = \frac{40}{3}\left( \partial_{\mu}\phi\partial_{\nu}\phi -\frac{1}{2}(\partial\phi)^2g_{\mu \nu}\right) \\
    &- \frac{1}{2}4 e^{-\frac{16}{3} \phi}\left(2L^8e^{-8\phi}-5\right)g_{\mu\nu}\,, \\
	\Box \phi & = 4e^{-\frac{16}{3} \phi}\left( 1 - L^8e^{-8\phi}\right)\,.
\end{split}
\end{align}
These equations follow from an effective 5d action for the 5d metric coupled to a scalar $\phi$ with potential $V$,
\begin{align}
\begin{split}
\label{E:S5}
    S_5& = \frac{\text{vol}(\mathbb{S}^5)}{2\kappa_{10}^2} \int d^5x\sqrt{-g} \left( R - \frac{40}{3}(\partial\phi)^2 - V(\phi)\right)\,, 
    \\
    V&(\phi) = 4e^{-\frac{16}{3}\phi} \left( 2 L^8 e^{-8\phi}-5\right)\,, 
\end{split}
\end{align}
which has a global minimum at $e^{\phi} = L$. In AdS/CFT the operator dual to $\phi$ is irrelevant with dimension $\Delta = 8$. This operator is known: it is $\text{tr}(F^4) + (\text{SUSY})$~\cite{Gubser:1998kv}. In $L=1$ units asymptotically AdS$_5\times\mathbb{S}^5$ regions have $\phi \to 0$ near the AdS boundary.

In a moment we also consider a generalization where the five-form flux varies in the noncompact directions with $L^4 = L^4(x)$. Such a variation is sourced by a density of D3-brane charge smeared homogeneously over the $\mathbb{S}^5$. The LHS of the 10d Einstein's equations continue to follow from variation with respect to~\eqref{E:S5}, although now there is an extra contribution to the RHS coming from the smeared 3-branes.

%--------------------------------------
\emph{Interfaces and boundaries.}~We would like to find $SO(6)$-invariant spacetimes with a boundary. If 10d IIB SUGRA had an action principle then we would simply deduce Neumann-like boundary conditions on the SUGRA fields by mandating a consistent variational principle. In the Appendix we are able to do just this for asymptotically AdS$_4\times\mathbb{S}^7$ and AdS$_7\times\mathbb{S}^4$ geometries in 11d SUGRA. Here, since such an action does not exist we work at the level of equations of motion and thus our approach is slightly roundabout. We first consider the dual of a conformal interface connecting a region with $N_+$ units of five-form flux to one with $N_-$ units where we take $N_+>N_-$. The total isometry is then $SO(2,3)\times SO(6)$. To get a boundary we will then take $N_-=0$ and cut off the geometry on that side of the interface in such a way as to respect the Neumann boundary condition on the metric that follows from pure 10d Einstein gravity, $K(H)_{MN} - K(H) H_{MN}=0$ with $H_{MN}$ the induced metric and $K(H)_{MN}$ its extrinsic curvature. To start we allow the five-form flux to vary in a continuous way corresponding to an $SO(2,3)\times SO(6)$-invariant density of 3-brane charge, i.e. we consider a ``fat'' interface. We will see that such a smooth interface is forbidden, and it must be ``sharp'' instead, with the 3-brane charge localized to a codimension-1 surface.

Let $\rho$ be a coordinate in the non-compact part of the geometry; the 5d geometry and five-form flux then read
\begin{align}
    \nonumber
    ds^2_5&= e^{2A(\rho)}ds^2_{\text{AdS}_4}+d\rho^2\,,
    \\
    \label{E:ansatz}
    F_{t x y z\rho} (\rho)&= \frac{4e^{-\frac{40}{3}\phi(\rho)+4A(\rho)}}{z^4}L^4(\rho)\,,
\end{align}
with $ds^2_{\text{AdS}_4} = \frac{-dt^2+dx^2+dy^2+dz^2}{z^2}$ and where $\frac{L^4}{4\pi \ell_s^4}$ interpolates between $ N_+$ at large positive $\rho$ and $  N_-$ at large negative $\rho$. 

From the divergence of the five-form flux we infer a density $\mathcal{J}^{txyz}$ of 3-brane charge,
\beq
	\frac{1}{(2\pi\ell_s)^4}D_MF^{Mtxyz} = -\frac{1}{4(\pi \ell_s)^4}\frac{\sqrt{\tilde{g}}}{\sqrt{-G}}\frac{\partial L^4}{\partial\rho}  = \mathcal{J}^{txyz}\,,
\eeq
where the indices are raised with the 10d Einstein metric. Integrating the 3-brane density over all of space and using $\text{vol}(\mathbb{S}^5)=\pi^3$ we find that the brane distribution carries $N_--N_+<0$ units of D3-brane charge as it ought.

Let us suppose that this distribution is made up out of a density of $\overline{\text{D}3}$-branes in the region where $\frac{\partial L^4}{\partial \rho}$ is positive and D3-branes where it is negative. The 10d stress tensor then receives a contribution from the 3-branes and the 5-form flux. It reads 
\begin{align}
	2\kappa_{10}^2T_{MN}dx^Mdx^N =& -\left|\frac{\partial L^4}{\partial \rho}\right|e^{-\frac{20\phi}{3}+2A} ds^2_{\text{AdS}_4}
	\\
	\nonumber
	& - 4 L^8 e^{-\frac{20\phi}{3}}ds_5^2+ 4 L^8 e^{-2\phi} d\Omega_5^2\,.
\end{align}
The conservation of the stress tensor implies
\beq
	 -6 e^{-\frac{20\phi}{3}}L^4 \frac{\partial L^4}{\partial\rho} + (3A' - 5 \phi')\left|\frac{\partial L^4}{\partial\rho}\right|= 0\,.
\eeq
In the region where $\frac{\partial L^4}{\partial \rho}>0$ one can show that the Einstein's equations imply
\beq
 	2 e^{-2A} + 8 e^{-\frac{40\phi}{3}} L^8 + e^{-\frac{20\phi}{3}} \frac{\partial L^4}{\partial \rho} = 0\,.
\eeq
However this is a sum of positive terms and thus it cannot vanish, leading to a contradiction. A similar argument applies in a region where $\frac{\partial L^4}{\partial \rho}<0$. Thus there cannot be a ``fat'' interface preserving $SO(2,3)\times SO(6)$ isometry. 

However a ``sharp'' interface localized at some value of $\rho$, call it $\bar{\rho}$, is allowed. For example, consider $L^4(\rho) = L^4 \text{sgn}(\rho-\bar{\rho})$ with $\frac{L^4}{\ell_s^4} = 4\pi N$, corresponding to $2N$ $\overline{\text{D}3}$ branes smeared along an interface at $\rho =\bar{\rho}$ separating a region $\rho>\bar{\rho}$ with $N$ units of five-form flux and one $\rho<\bar{\rho}$ with $-N$ units. Integrating the angular and AdS$_4$ components of the 10d Einstein's equations in a pillbox surrounding $\rho=\bar{\rho}$ we find that the changes in $A'$ and $\phi'$ are fixed as
\begin{align}
\begin{split}
	3\Delta A' & = 2\Delta \phi'\,,
	\\
	\Delta A' & = - \frac{2L^4}{3}e^{-\frac{20\phi}{3}}\,,
\end{split}
\end{align}
or, covariantly,
\begin{align}
\begin{split}
\label{E:Israel}
	\Delta K(H)_{\mu\nu}-\Delta K(H) H_{\mu\nu}&= -4L^4 e^{-5\phi} H_{\mu\nu}\,,
	\\
	\Delta K(H)_{\alpha\beta}-\Delta K(H)H_{\alpha \beta} &= 0\,,
\end{split}
\end{align}
in terms of the induced metric $H_{MN}$ at $\rho = \bar{\rho}$ and the change in its extrinsic curvature $\Delta K(H)_{MN}$. These are the junction conditions for such a geometry. The RHS of the first line is the stress-energy sourced by the smeared 3-branes. As we will see, the angular condition is equivalent to a statement of force balance on the 3-branes.

We can find an asymptotically AdS$_5\times\mathbb{S}^5$ solution to the 10d Einstein's equations subject to this junction condition. It is convenient to group the 10d Einstein's equations into the equation for the sphere warpfactor $\phi$, the 5d Einstein's equation contracted with the null vector $u^{\mu}\partial_{\mu} =z e^{-A}\partial_t + \partial_{\rho}$, and the $\rho\rho$ component of the 5d Einstein's equations. The former two read in $L=1$ units
\begin{align}
\begin{split}
\label{E:eoms}
    \phi''+4A' \phi' &= 4e^{-\frac{16}{3} \phi}\left( 1 - e^{-8\phi}\right)\,, \\
    A'' - e^{-2A} &= -\frac{40}{9}\phi'^2\,,
\end{split}
\end{align}
and the last is a constraint that, on either side of the interface, fixes $A'$ in terms of $\{\phi,A,\phi'\}$. Fixing a reflection symmetry whereby $A$ and $\phi$ are continuous and even functions about the interface, the $\rho\rho$ component imposes 
\beq
\label{E:rhoConstraint}
	e^{-2A(\bar{\rho})} = \frac{5}{3}e^{-\frac{16}{3}\phi(\bar{\rho})}\,.
\eeq
In this way the data $\{A,\phi',A'\}$ on either side of the interface is fixed in terms of $\phi$ there, which we regard as an initial condition when solving the equations~\eqref{E:eoms} numerically. We tune this initial condition so that the 5d geometry is asymptotically AdS$_5$ at large $\rho$, which requires $\rho\to 0$.

To find a solution describing a boundary we can either (1) orbifold across the interface, restricting ourself to the portion of the geometry with $\rho>\bar{\rho}$, or (2) instead consider an interface connecting a region with $N$ units of flux for $\rho>\bar{\rho}$  to one with no flux for $\rho<\bar{\rho}$, with the further condition that there is zero extrinsic curvature as $\rho\to_+ \bar{\rho}$ so that we can consistently excise the geometry in that region. In the latter the 10d junction conditions read in $L=1$ units
\begin{align}
\begin{split}
\label{E:Israel2}
	K(H)_{\mu\nu}-K(H) H_{\mu\nu}&= -2 e^{-5\phi} H_{\mu\nu}\,,
	\\
	 K(H)_{\alpha\beta}-K(H)H_{\alpha \beta} &= 0\,,
\end{split}
\end{align}
for $K(H)_{MN}$ the extrinsic curvature of the slice $\rho\to_+\bar{\rho}$. In terms of the warpfactors $A$ and $\phi$, these read
\beq
	\phi'(\bar{\rho}) = - e^{-\frac{20}{3}\phi(\bar{\rho})}\,, \quad A'(\bar{\rho}) = -\frac{2}{3}e^{-\frac{20}{3}\phi(\bar{\rho})}\,,
\eeq
and the $\rho\rho$ component of the 5d Einstein's equations fixes the constraint~\eqref{E:rhoConstraint} on the data at the boundary. In fact these are equivalent to the boundary conditions~\eqref{E:Israel} for a symmetric interface discussed above. As a result integrating the 10d Einstein's equations away from the boundary leads to the $\rho>\bar{\rho}$ portion of a symmetric interface. Equivalently, a symmetric interface is a doubled version of the problem with a boundary.

\begin{figure}[t]
\begin{center}
\includegraphics[width=3in]{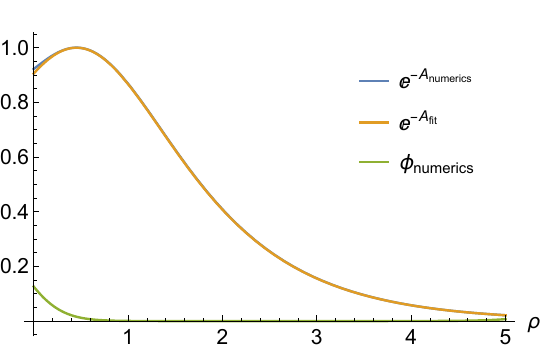} 
\caption{\label{F:AdS5} The profiles for $\phi$ (green) and $e^{-A}$ (blue) in our numerical solution compared to a fit (orange) of $e^{-A}$ to $\frac{1}{\cosh(\rho-\rho_0)}$ where with the fit $\rho_0\approx 0.453$. Because this fit is very good our solution is well-approximated by a simple model of AdS$_5$ spacetime ending on a tensionful brane of tension $T\approx 1.27$.}
\end{center}
\end{figure}

%--------------------------------------
\emph{Solution.}~We found a geometry with a smeared ETW brane by numerically solving~\eqref{E:eoms}, setting $\bar{\rho}=0$ as a choice of coordinates. This solution has an asymptotically AdS$_5\times\mathbb{S}^5$ region at large $e^\rho$, which we present in Fig.~\ref{F:AdS5}. We are unable to tune the initial conditions so that we exactly achieve the desired boundary condition $\lim_{\rho\to\infty}\phi = 0$. For $\rho\gtrsim 5$, $\phi$ begins to grow exponentially, as one would expect since it is dual to an irrelevant operator and the growing mode can be interpreted as a source for the dual operator which requires exponentially good fine-tuning to set to vanish. 

For $\rho \lesssim 5$, however, we find that our solution is very well-approximated by pure AdS$_5$ cut off by a tensionful brane with tension $T\approx 1.27$, as plotted in Fig.~\ref{F:AdS5}. The deviations from pure AdS$_5$ are concentrated in the part of the spacetime very close to the ETW brane where $\phi$ is appreciable, and even then the deviations in $e^{-A}$ are only at the percent level.

An important property of holographic ETW branes is that they are connected to the boundary by null geodesics that traverse a finite amount of boundary time when the dual BCFT is placed on a hemisphere. This leads to singularities in two-point functions of the dual BCFT at cross-ratios other than those mandated by operator-product-expansion limits~\cite{Reeves:2021sab}. As we show in~\cite{constraints} and discuss in the Appendix these null geodesics are characterized by the quantity
\begin{equation}
	\phi_b = \int_{\bar{\rho}}^{\infty} d\rho \,e^{-A(\rho)} \,.
\end{equation}
Numerically evaluating this quantity we find $\phi_b\approx 2.01 < \pi$. By computing $\phi_b$ in simple bottom-up models of ETW branes, we find that our configuration can be modeled by a tensionful brane coupled to Einstein gravity with an effective tension $T\approx 1.27$.

We now study the stability of the smeared ETW brane. Consider the effective action for a single $\overline{\text{D}3}$ brane in the distribution, 
\begin{equation}
    S = - \frac{1}{(2\pi)^3\ell_s^4}\left( \int d^4\sigma \sqrt{-\text{P}[g]}- \int C_4\right)\,,
\end{equation}
located at some point on the $\mathbb{S}^5$. When it sits at $\rho=\bar{\rho}$ it is extended along an AdS$_4$ submanifold inside of the five-dimensional part of the geometry, and its intersection with the conformal boundary of the space corresponds to the boundary of the dual BCFT. 

\begin{figure}[t!]
\begin{center}
\includegraphics[width=2.8in]{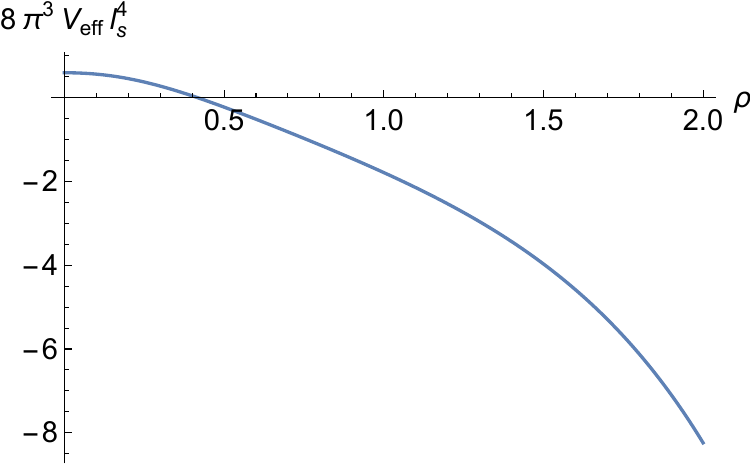}
\caption{\label{F:potential} The effective potential for a single 3-brane as a function of its radial position $\rho$. The potential is globally maximized at $\rho=\bar{\rho}=0$, indicating an instability to separating from the smeared ETW brane.}
\end{center}
\end{figure}

Now suppose the brane sits at some other constant value of $\rho$, in which case it still ends on the boundary of the dual BCFT. The effective potential for $\rho$ is
\begin{equation}
    V_{\rm eff} \propto e^{4A-\frac{20}{3}\phi}- 4\int_{\bar{\rho}}^{\rho}d\rho' e^{4A(\rho')-\frac{40}{3}\phi(\rho')}\,,
\end{equation}
where the first term comes from the tension part of the brane action and the second from the Wess-Zumino term. We plot this effective potential in Fig.~\ref{F:potential}. It is extremized at $\rho=\bar{\rho}$ implying that there is force balance on a $\overline{\text{D}3}$ brane there. It is easy to show that this is equivalent to the angular part of the Israel junction condition~\eqref{E:Israel2}. However, the potential is not minimized at $\rho=\bar{\rho}$, but instead globally maximized. In fact the most energetically favorable possibility is that this single 3-brane separates from the rest of its distribution and tends towards the conformal boundary as $\rho\to\infty$. This is an instability.

%--------------------------------------
\emph{Discussion}.~In this Letter we endeavoured to embed the simple model of~\cite{Takayanagi:2011zk} for ETW branes in holography into the lamppost examples of AdS/CFT. We mandated that our supergravity backgrounds were invariant under rotations of the ``internal space,'' the $\mathbb{S}^5$ for the AdS$_5\times\mathbb{S}^5$ vacuum of type IIB SUGRA, with an ETW brane dual to a conformal boundary of the dual CFT. Because these vacua are supported by RR fluxes, imposing a Neumann condition on the RR fields implies that the ETW brane carries RR charge, and in particular is a smeared distribution of branes. In the body of this Letter we presented a solution whereby an asymptotically AdS$_5\times\mathbb{S}^5$ solution of type IIB SUGRA ends on a smeared distribution of 3-branes in such a way as to respect the appropriate analogue of the Israel junction condition. The ensuing geometry is quite nearly AdS$_5\times\mathbb{S}^5$ for most of the spacetime up to a region very close to the smeared ETW brane. However, we found an instability whereby the individual 3-branes making up the ETW brane are unstable to the fluctuations of their transverse scalars. 

Normally one does not draw too much attention to an unstable solution in SUGRA. Here we are doing the opposite, because the ingredients of our construction, in particular the treatment of the RR fields, will appear in any attempt to embed this simple model into AdS/CFT when mandating isometry of the internal space.

While this solution is very well approximated by AdS$_5\times\mathbb{S}^5$ cut off by a tensionful brane wrapping the $\mathbb{S}^5$, we stress that all of the details of type IIB SUGRA, including the effective potential for the $\mathbb{S}^5$ warpfactor and the angular part of the Israel junction condition, are all important in finding it. In particular, the angular part of the junction condition expresses the statement that the 3-branes at the end of the spacetime are in equilibrium, with the gravitational and RR forces on them balancing.

We also adapted our methods to search for $SO(6)$-invariant domain walls that connect two asymptotically AdS$_5\times\mathbb{S}^5$ regions, one supported by $N_+$ units of five-form flux and the other by $N_-$ units. For $N_+=-N_-$ the interface geometry is a doubled version of the one with a boundary. However, for general $N_-/N_+$ we do not find solutions that respect the boundary conditions far away from the interface. We can tune initial conditions at an interface to achieve an asymptotically AdS$_5\times\mathbb{S}^5$ region on one side of the interface, or the other, but not both. 

Let us present a candidate for the (evidently, unstable) boundary condition dual to this unstable ETW brane. Since the ETW brane carries $N$ units of 3-brane charge, we have $N$ ``color'' branes ending on $N$ ``ETW'' branes. Let us for the moment take $N=1$. Our bulk solution suggests that this configuration would involve a single color brane and a single ETW brane in 10d flat space. In this suggestion there is a color brane extended in the $0123$ directions of flat space, ending on an ETW brane extended along the $0124$ directions at $x^3=0$. Both the color and ETW branes are semi-infinite, the former extended along $x^3>0$ and the latter along $x^4>0$. At small string coupling the semi-infinite color and ETW branes join into a single 3-brane. This is a non-supersymmetric, unstable ``2 ND'' intersection for which the boundary conditions for the fields on the color brane are known. In particular the scalar $\phi^4$ obeys a Neumann condition at $x^3=0$ but the remaining five scalars $\phi^{5,..,9}$ obey a Dirichlet condition. This configuration thus breaks the $SO(6)$ R-symmetry down to the $SO(5)$ that rotates the 56789 directions. For $N>1$ we can now consider $N$ semi-infinite color branes ending on $N$ semi-infinite ETW branes, with a similar $SO(6)\to SO(5)$ breaking boundary condition for each of the color branes, but now we can choose different $SO(5)$ subgroups for each. Such an arrangement will at finite $N$ completely break the $SO(6)$ R-symmetry. Our candidate dual to our bulk solution is precisely this in the $N\to\infty$ limit where the $SO(6)$ symmetry is restored by a uniform distribution of $SO(6)\to SO(5)$ breaking boundary conditions for each of the color branes.

From this point of view the instability of our ETW brane is perhaps not so surprising. However we stress that one necessarily confronts these 2 ND intersections when taking the model of~\cite{Takayanagi:2011zk} seriously and attempting to embed it into string theory without using orientifolds.

The main lesson of our analysis is the importance of embedding models of ETW branes into string theory, especially from the point of view of checking stability.

\emph{ Acknowledgments.}
We would like to thank A.~Karch, D.~Neuenfeld, M.~Rozali, J.~Sorce, and J.~Sully for enlightening discussions. This work was supported in part by an NSERC Discovery Grant.

\bibliographystyle{apsrev4-1}
\bibliography{refs}

\appendix

%----------------------------------------
\section{Light crossing and ETW branes}
%----------------------------------------

Consider the holographic dual of a BCFT, where the bulk spacetime has a line element
\begin{equation}
\label{E:generalBCFTbackground}
    ds^2=d\rho^2+e^{2A(\rho)}\left(\frac{-dt^2+d\Vec{x}^2+dz^2}{z^2}\right)\,,
\end{equation}
for $\rho \geq \bar{\rho}$ with an ETW brane located at $\rho = \bar{\rho}$. In particular, consider the gravity dual of a two-point function of BCFT operators $\langle O(t_1,\vec{x}_1,z_1)O(t_2,\vec{x}_2,z_2)\rangle$. Owing to the existence of the boundary, conformal symmetry constrains this correlation function to be a kinematical prefactor times a function of a single conformally-invariant cross-ratio formed from $x_1$ and $x_2$. The authors of~\cite{Reeves:2021sab} have argued that these two-point functions  have ``unphysical'' singularities, meaning singularities at cross-ratios that do not correspond to OPE limits either of the two $O$'s with each other, or an $O$ with the boundary, corresponding in the bulk to null geodesics that connect $x_1$ and $x_2$ by bouncing off the ETW brane. They show that this is the case both in the simple model~\cite{Takayanagi:2011zk} of spacetime ending on a tensionful brane as well as for a stringy realization of an interface CFT, namely $\mathcal{N}=4$ super-Yang Mills subject to a supersymmetric Janus deformation. More generally one expects this to be true by the standard geometric optics argument for large dimension boundary operators, or by rewriting the boundary-to-boundary propagator as a sum over worldlines. The ensuing ``unphysical'' singularity is the AdS/BCFT version of the ``looking for a bulk point'' singularity of~\cite{Maldacena:2015iua}.

\begin{figure}[t]
\begin{center}
\includegraphics[width=3in]{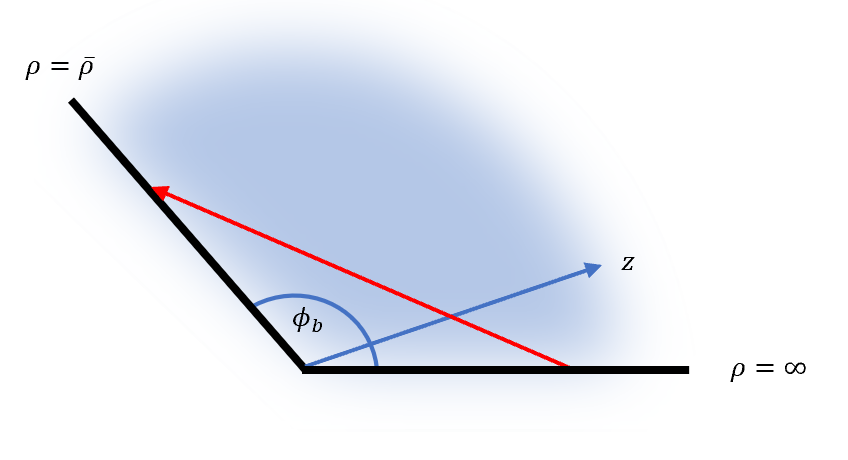} 
\end{center}
\caption{Null ray (red) traveling from the AdS boundary ($\rho=\infty$) to an ETW brane ($\rho=\bar{\rho}$) for $\phi_b<\pi$}
\end{figure}

Performing a coordinate transformation $d\phi=-e^{-A}d\rho$, the line element~\eqref{E:generalBCFTbackground} becomes
\begin{equation}
    ds^2=\frac{e^{2A(\rho)}}{z^2}\left(-dt^2+d\Vec{x}^2+dz^2+z^2d\phi^2\right)\,.
\end{equation}
From the coordinate transformation, we have 
\begin{equation}\label{LightCrossingTime}
    \phi=\int_\rho^\infty e^{-A(\rho')}d\rho' \,. 
\end{equation}
In this coordinate system, $\rho=\infty$ corresponds to $\phi=0$ and we define $\phi_b$ to be the location of ETW brane at $\rho=\bar{\rho}$. We note that $-dt^2+d\Vec{x}^2+dz^2 + z^2 d\phi^2$ is just a wedge of flat space written as $\mathbb{R}^{d-1}$ times a wedge of $\mathbb{R}^2$ described in polar coordinates with $z$ the radius and $\phi$ the angle. Null geodesics in this space are just straight lines. One can always boost so that such a geodesic sits at constant $\vec{x}$, so that the motion takes place in $(t,z,\phi)$. Clearly, a null ray starting from AdS boundary cannot travel to the ETW brane if $\phi_b\geq\pi$. This is true when the bulk geometry is sliced by the Poincar\'e patch of AdS. However, if instead we take global slices, so that the dual BCFT is formulated on a hemisphere, then these geodesics always exist and, if the two boundary points connected by the trajectory are at the pole of the hemisphere, then $\phi_b$ is the global time of travel to travel to the ETW brane.

%----------------------------------------
\section{Smeared ETW branes in 11d SUGRA}
%----------------------------------------
In this Appendix we uncover realizations of smeared ETW branes in 11d SUGRA, in the context of the AdS$_4\times\mathbb{S}^7$ and AdS$_7\times\mathbb{S}^4$ vacua of M-theory. The derivation of these backgrounds closely follows that presented in the main text for the $SO(6)$-invariant compactification of type IIB SUGRA.

We begin with a brief primer to 11d supergravity, whose bosonic fields are a metric $G_{\mu \nu}$ and a Ramond-Ramond potential $C_{3}$. The bosonic action is given by
\begin{align}
\begin{split}
    S_{11} = &\frac{1}{2 \kappa^{2}_{11}} \int d^{11}x \sqrt{-G} \left(R - \frac{1}{2} |F_{4}|^{2} \right) 
    \\
    & \qquad\qquad  - \frac{1}{12\kappa_{11}^2} \int C_{3} \wedge F_{4} \wedge F_{4}\,,
\end{split}
\end{align}
and the field equations are
\begin{widetext}
\begin{align}
\begin{split}
    R_{MN} - \frac{R}{2} G_{MN} &= \frac{1}{2} F_{MPQR}F_{N}^{PQR} - \frac{1}{96} G_{MN} F_{PQRS}F^{PQRS}\,, \\
    D_{M}F^{MNPQ} &= \frac{1}{2 \times 4!^{2}} \epsilon^{NPQS_{1}...S_{4}T_{1}...T_{4}}F_{S_{1}...S_{4}}F_{T_{1}...T_{4}}\,,
\end{split}
\end{align}
\end{widetext}
where $\epsilon^{NPQS_{1}...S_{4,7}T_{1}...T_{4,7}}$ is the 11d Levi-Civita tensor. These equations have AdS$_4\times\mathbb{S}^7$ and AdS$_7\times\mathbb{S}^4$ solutions respectively supported by a stack of M2 or M5 branes.

%----------------------------------------
\subsection{Asymptotically AdS$_{4} \times \mathbb{S}^{7}$ geometries}
%----------------------------------------

We begin with the $SO(8)$ invariant compactification of 11d SUGRA on an $\mathbb{S}^7$, with $N$ units of flux through it. We parameterize the most general $SO(8)$-invariant line element and conserved four-form flux as
\begin{equation}\label{AdS4S7Metric}
\begin{split}
        ds_{11}^{2} &= e^{-7 \phi} g_{\mu \nu} dx^{\mu}dx^{\nu} + e^{2 \phi} d \Omega^{2}_{4}\,, \\ 
        F_{\mu\nu\rho\sigma} &= 6 L^{6}e^{-21\phi} \varepsilon_{\mu\nu\rho\sigma}\,, 
\end{split}
\end{equation}
where $g_{\mu\nu}$ is a four-dimensional metric and $\epsilon_{\mu\nu\rho\sigma}$ its corresponding Levi-Civita tensor. This ansatz automatically satisfies the RR three-form equations of motion. The four-dimensional part of the 11d line element is scaled by the warpfactor $\phi (\rho)$ as a good choice of Weyl shift such that, upon dimensional reduction, the action is in Einstein-Hilbert form. Indeed, after dimensional reduction on the $\mathbb{S}^{7}$ the action reads
\begin{equation}\label{dimreducedS7action}
    S_4 = \frac{\text{vol}(\mathbb{S}^{7})}{2 \kappa^{2}_{11}} \int d^{4}x \sqrt{-g} \left(R  - \frac{63}{2} (\partial \phi)^{2} - V(\phi)  \right) + (\text{bdy})\,,
\end{equation}
where $R$ is the 4d scalar curvature
%, and the warp factor $\phi (\rho)$ behaves as the matter content. The potential
and the potential is given by
\begin{equation}
    V(\phi) = 6 e^{-9 \phi} \left(3 L^{12} e^{-12 \phi} - 7 \right)\,,
\end{equation}
with a unique minimum at $e^{\phi} = L$. Freezing $\phi$ to that value gives the Einstein-Hilbert action with negative cosmological constant
\begin{equation}
    S_4 = \frac{\text{vol}(\mathbb{S}^{7})}{2 \kappa^{2}_{11}} \int d^{4}x \sqrt{-g} \left(R + \frac{24}{L^{9}} \right) + (\text{bdy})\,,
\end{equation}
for $\Lambda = -\frac{12}{L^{9}} = -\frac{3}{L^{2}_{\rm eff}}$. We henceforth choose $L=1$ as a choice of units so that the effective AdS$_4$ radius is $L_{\rm eff} = \frac{1}{2}$. Small fluctuations of $\phi$ around this minimum, $\phi = \delta \phi$, are described by the quadratic action
\begin{equation}
    S_{\rm quad} = -\frac{63 \, \text{vol}(\mathbb{S}^{7})}{2 \kappa^{2}_{11}} \int d^{4}x \sqrt{-g} \left(\frac{1}{2} (\partial \delta \phi)^{2} + \frac{9}{L^{2}_{\rm eff}} \delta \phi^{2} \right)\,.
\end{equation}
These fluctuations have an effective mass-squared $m^{2} L^{2}_{\rm eff} = 18$ so that the dual operator is dimension $\Delta = 6$ and so irrelevant. Asymptotically AdS$_4\times\mathbb{S}^7$ regions thus have $\phi \to 0$.
For reference the equations of motion for $g_{\mu\nu}$ and $\phi$ are
\begin{align}\label{GeneralEOMSAdS7S4}
     R_{\mu \nu} - \frac{R}{2}g_{\mu \nu} &= \frac{63}{2} \left(\partial_{\mu} \phi \partial_{\nu} \phi - \frac{1}{2} (\partial \phi)^{2}g_{\mu\nu} \right)  
     \\
     \nonumber
     &\qquad \quad- \frac{1}{2} 6 e^{-9 \phi} \left(3 L^{12} e^{-12 \phi} - 7 \right) g_{\mu \nu}\,, 
     \\
     \nonumber
     \square \phi& = 6e^{-9 \phi} \left( 1 - L^{12}e^{-12 \phi} \right)\,.
\end{align}

%----------------------------------------
\subsubsection{ETW brane}
%----------------------------------------

Mirroring the discussion in the main text we endeavor to put an $SO(8)$-invariant ETW brane into this geometry. One way to do this is along the lines of our analysis in the main text, namely to consider interfaces that connect a region with $N_+$ units of flux to one with $N_-$ units, and then take $N_-\to 0$. However there is a simpler option available to us because, unlike in IIB supergravity, we have an action principle for 11d SUGRA, and so we can enforce a consistent variational principle for the SUGRA fields, which will in turn imply that the boundary carries RR charge.

As in our IIB analysis we study backgrounds that preserve the isometry of the internal space. Because we also have in mind the dual of a conformal boundary condition, we then impose that the bulk geometry has a $SO(2,2)\times SO(8)$ isometry, so that we can parameterize the 4d line element as
\begin{equation}
ds^{2} = e^{2 A(\rho)} \left( \frac{-dt^2 +dx^2+ dz^2}{z^2} \right)+d\rho^2\,,
\end{equation}
where the warpfactor $A$ is fixed to be a function of $\rho$ alone, and we take the spacetime to exist only for $\rho\geq \bar{\rho}$. With this parameterization we have
\begin{equation}
    F_{txz \rho} = \frac{6 e^{-21 \phi + 3 A}}{z^{3}}.
\end{equation}
We proceed to find boundary conditions at $\rho=\bar{\rho}$ such that, with suitable boundary terms, the SUGRA effective action has a consistent variational principle. In backgrounds where $F_4\wedge F_4 = 0$ the variation of the RR field has the boundary term
\beq
\label{E:delta11bdy}
    \delta S_{11} = -\frac{1}{2\kappa_{11}^2\times 3!}\int d^{10}x \sqrt{-H}\,\delta C_{MNP}N_QF^{Q MNP}\,,
\eeq
where indices are raised with the 11-dimensional metric and $N_Q$ is a normalized outward-pointing unit vector. We would like to impose a Neumann-like condition on $C_3$, but this cannot be achieved without a RR current density on the boundary, since at $\rho=\bar{\rho}$ we have $\sqrt{-H}N_Q F^{Qtxz} = - 6\sqrt{\tilde{g}}$ where $\tilde{g}$ is the metric on a unit $\mathbb{S}^7$. Allowing a RR charge density $J^{txz}$ at $\rho=\bar{\rho}$ introduces an extra boundary term in the variation of the action, so that
\beq
    \delta S_{11} = \int d^{10}x \sqrt{-g} \,\delta C_{txz} \left( \frac{1}{2\kappa_{11}^2}F^{\rho txz} - J^{txz}\right)\,.
\eeq
We achieve a Neumann-like condition by setting the term in parentheses to vanish, which implies a uniform density of $N$ units of M2-brane charge smeared homogeneously over the $\mathbb{S}^7$. We then assume that this distribution of charge is built from $N$ M2 branes smeared over the boundary, which leads to a contribution to the stresse-energy there. Including the Gibbons-Hawking term as usual, we also impose a Neumann-like condition on the metric, the analogue of the Israel junction condition, $K(H)_{MN}-K(H)H_{MN}=-T_{MN}$,which guarantees that the boundary term in the variation of the action with respect to metric fluctuations vanishes.  If we let $\Sigma = \text{vol}_{\mathbb{S}^7}=d^7\theta \,\sqrt{\tilde{g}}$ denote the smearing form, then the total boundary term, the sum of the Gibbons-Hawking term and the action of this distribution of $\text{M}2$-branes, reads
\begin{align}
\label{E:M2ETW}
    S_{\rm bdy}& = \frac{1}{2 \kappa_{11}^2} \bigg(2 \int d^{10} x \sqrt{-H}\, K(H) 
    \\
    \nonumber
    &-6 \left(\int d^{3} \sigma \, \sqrt{-\text{P}[G]} \wedge \Sigma + \int C_{3} \wedge \Sigma \right) \bigg)\,,
\end{align}
where $\Sigma = \text{vol}_{\mathbb{S}^7} = d^7\theta \sqrt{\tilde{g}}$ is the smearing form and $H$ is the induced metric on the boundary.

The analogue of the Israel Junction condition is
\begin{align}
\begin{split}
\label{A4S7IJCondition}
     K(H)_{\mu \nu} - K(H) H_{\mu \nu} &= -3 e^{-7\phi} H_{\mu \nu}\,, 
     \\ 
    K(H)_{\alpha \beta} - K(H) H_{\alpha \beta} &= 0\,.
\end{split}
\end{align}
Upon using the trace of the first line, the angular part can be shown to be equivalent to the statement that the distribution of M2 branes are in equilibrium. Expressed in terms of $A$ and $\phi$ the junction condition simplifies to
\begin{align}
\begin{split}
    \label{A4S7BCs}
    A' & = -\frac{3}{2} e^{-\frac{21}{2}\phi}\,,
    \\
    \phi' & = -e^{-\frac{21}{2}\phi}\,.
       %K(h)_{\mu \nu} - K(h) h_{\mu \nu} & = -3 e^{-14 \phi} h_{\mu \nu} \,,
       %\\
       %  K(h)  + \frac{9}{2} \partial_{\rho} \phi & = 0\,,
\end{split}
\end{align}
at $\rho = \bar{\rho}$.

We may also consider interfaces connecting two asymptotically AdS$_4\times\mathbb{S}^7$ regions in a $SO(2,2)\times SO(8)$ invariant way. As in IIB SUGRA, such interfaces must be sharp so as to be consistent with the Einstein's equations. Let the region with $\rho>\bar{\rho}$ be supported by $N_+$ units of four-form flux, and the region with $\rho<\bar{\rho}$ be supported with $N_-$ units. The flux background is $F_{\mu\nu\rho\sigma} = 6 (L_{+}^6 \Theta(\rho-\bar{\rho}) + L_-^6 \Theta(\bar{\rho}-\rho))e^{-21\phi}\varepsilon_{\mu\nu\rho\sigma}$ where $\frac{L_{\pm}^6}{\ell_{P}^2} = 32\pi^2 N_{\pm}$ with $\ell_P$ the 11-dimensional Planck length. This flux background is not conserved, implying the existence of $N_+-N_-$ units of M2-brane charge at $\rho=\bar{\rho}$, smeared homogeneously over the $\mathbb{S}^7$. Assuming that this charge density is made up by $N_+-N_-$ M2 branes, and their concomitant stress energy, one ends up with 11-dimensional SUGRA together with this brane source at $\rho=\bar{\rho}$. For $N_+>N_-$ we have 
\begin{align}
\label{E:M2interface}
    S &= S_{11} + S_{\rm interface}\,,
    \\
    \nonumber
    S_{\rm interface} & = -\frac{N_+-N_-}{\text{vol}(\mathbb{S}^7)}T_3 \left( \int d^3\sigma \sqrt{\-\text{P}[G]}\wedge \Sigma + \int C_3 \wedge \Sigma\right)\,,
\end{align}
where $T_3 = \frac{1}{(2\pi)^2\ell_P^3}$ is the tension of a single M2 brane. There is an Israel junction condition at $\rho=\bar{\rho}$ that comes from integrating the Einstein's equations in a pillbox around $\rho=\bar{\rho}$. The $N_-\to 0$ limit of that junction condition is simply the one we reported above in~\eqref{A4S7IJCondition}; the $N_-\to 0$ limit of the interface action also reduces to the M2-brane part of the ETW action in~\eqref{E:M2ETW}.

%----------------------------------------
\subsubsection{Solution}
%----------------------------------------

We find it convenient to numerically solve the scalar equation along with the component of the Einstein's equations in~\eqref{GeneralEOMSAdS7S4} obtained after contraction with the null vector
$u^{\mu} \partial_{\mu} = z e^{-A}\partial_{t} + \partial_{\rho}$. These are 
\begin{align}
\begin{split}
\label{AdS4EOMs}
    A'' -e^{-2 A} &= -\frac{63}{4} \phi'^{2}\,, 
    \\
     \phi'' + 3 A' \phi' &= 6 e^{-9 \phi} \left( 1 - e^{-12 \phi} \right)\,.
\end{split}
\end{align}
Using the Israel junction condition~\eqref{A4S7BCs}, the $\rho\rho$-component of the Einstein's equations imposes a further constraint on the data at $\rho=\bar{\rho}$,
\begin{equation}\label{A4S7ExtraBC}
    e^{-2 A} = 7 e^{-9 \phi }\,.
\end{equation}
This, together with the junction condition, determines $(A,\phi',A')$ in terms of $\phi$ there.

\begin{figure}[t]
\begin{center}
\includegraphics[width=3in]{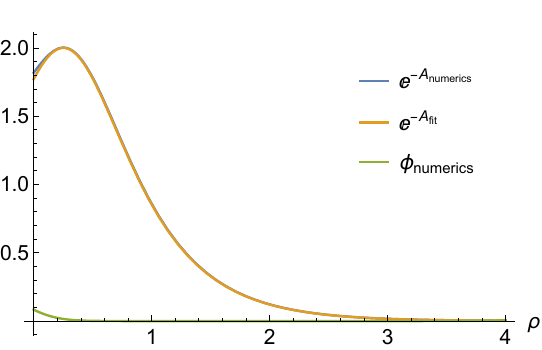} 
\end{center}
\caption{\label{F:AdS4} The numerical profiles for $\phi(\rho)$ (green) and $e^{-A(\rho)}$ (blue) for an asymptotically AdS$_4\times\mathbb{S}^7$ geometry ending on a distribution of M2 branes at $\rho=0$. The 4d geometry is very close to that of empty AdS$_4$ ending on a tensionful brane of tension $T\approx 1.89$, as indicated by the fit (orange).}
\end{figure}

Our strategy is to solve the equations~\eqref{AdS4EOMs} by shooting from $\rho=\bar{\rho}$ and dialing the ``initial condition'' $\phi(\bar{\rho})$ until we achieve the asymptotically AdS$_4\times\mathbb{S}^7$ boundary condition $\phi\to 0$ as $\rho\to\infty$.
We are unable to impose this boundary condition exactly, but we have found a solution which obeys it up to a fairly large value of $\rho \lesssim 4$. The solution is plotted in Fig.~\ref{F:AdS4}. The 4d part of the solution is very close to pure AdS$_4$ cut off by a tensionful brane of tension $T\approx 1.89$. 

We have discussed the quantity $\phi_b$, which when $<\pi$ indicates that the ETW brane can be reached by a null geodesic from the AdS boundary. This solution has
\begin{equation}
    \phi_{b} = \int_0^{\infty} d \rho\, e^{-A(\rho)} \approx 2.06 < \pi\,.
\end{equation}
We also consider the effective potential for a single M2 brane separated from the distribution at $\rho=\bar{\rho}=0$, but extended along an AdS$_3$ slice at constant $\rho$ and at a fixed angle on the $\mathbb{S}^7$. Its effective potential obeys
\begin{equation}
    V_{\rm eff} \propto e^{3A-\frac{21}{2}\phi } - 6\int^{\rho}_0 d\rho' \,e^{3A(\rho') -21\phi(\rho')}\,,
\end{equation}
and it is globally maximized at $\rho=0$. As for our smeared distribution of 3-branes, this indicates an instability to fragmentation.

%----------------------------------------
\subsection{Asymptotically AdS$_{7} \times \mathbb{S}^{4}$ geometries}
%----------------------------------------

Now consider the $SO(5)$-invariant compactification on a $\mathbb{S}^4$ with $N$ units of flux through it. The most general line element and conserved four-form flux consistent with $SO(5)$ isometry are
\begin{align}
\begin{split}
\label{AdS7S4Metric}
        ds^{2} &= e^{-\frac{8}{5} \phi} g_{\mu \nu} dx^{\mu}dx^{\nu} + e^{2 \phi} d \Omega^{2}_{4}\,, 
        \\ 
        F_{\alpha\beta\gamma\delta} &= 3L^3 \varepsilon_{\alpha\beta\gamma\delta}\,,
\end{split}
\end{align}
where $\varepsilon_{\alpha\beta\gamma\delta}$ is the Levi-Civita tensor on a unit $\mathbb{S}^4$. The flux 
satisfies the RR equations of motion. The seven-dimensional effective action is
\begin{equation}\label{dimreducedS4action}
    S_7 = \frac{\text{vol}(\mathbb{S}^{4})}{2 \kappa^{2}_{11}} \int d^{7}x \sqrt{-g} \left(R  - \frac{36}{5} (\partial \phi)^{2} - V(\phi)  \right) + (\text{bdy})\,,
\end{equation}
where $R$ is the 7d scalar curvature and the potential is
\begin{equation}
    V(\phi) = 3 e^{-\frac{18}{5} \phi} \left(\frac{3 L^{6}}{2} e^{-6 \phi} - 4 \right)\,,
\end{equation}
which has a unique minimum at $e^{\phi} = L$. Freezing $\phi$ to that value gives the Einstein-Hilbert action with negative cosmological constant
\begin{equation}
    S_7 = \frac{\text{vol}(\mathbb{S}^{4})}{2 \kappa^{2}_{11}} \int d^{7}x \sqrt{-g} \left(R + \frac{15}{2L^{\frac{18}{5}}} \right) + (\text{bdy})\,,
\end{equation}
i.e. $\Lambda = -\frac{15}{4 L^{\frac{18}{5}}} = -\frac{15}{L^{2}_{\rm eff}}$. We henceforth work in $L=1$ units for which $L_{\rm eff} = 2$. Small fluctuations of $\phi$ around the minimum are described by a quadratic action
\begin{equation}
    S_{\rm quad} = -\frac{72 \, \text{vol}(\mathbb{S}^{4})}{10 \kappa^{2}_{11}} \int d^{7}x \sqrt{-g} \left(\frac{1}{2} (\partial \delta \phi)^{2} + \frac{36}{L^{2}_{\rm eff}} \delta \phi^{2} \right)\,.
\end{equation}
These fluctuations have an effective mass-squared $m^{2} L^{2}_{\rm eff} = 72$, and so the dual operator is irrelevant with $\Delta = 12$. Asymptotically AdS$_7\times\mathbb{S}^4$ geometries then have $\phi \to 0$. For reference the equations of motion are
\begin{align}
\label{GeneralEOMSAdS7S4}
     R_{\mu \nu} - \frac{R}{2}g_{\mu \nu} &= \frac{36}{5} \left(\partial_{\mu} \phi \partial_{\nu} \phi - \frac{1}{2} (\partial \phi)^{2}g_{\mu\nu} \right)  
     \\
     \nonumber
     & \qquad - \frac{3}{2} e^{-\frac{18}{5} \phi} \left(\frac{3L^6}{2} e^{-6 \phi} - 4 \right) g_{\mu \nu}\,, 
     \\
     \nonumber
     \square \phi &= 3e^{-\frac{18}{5} \phi} \left( 1 - L^{6}e^{-6 \phi} \right)\,.
\end{align}

%----------------------------------------
\subsubsection{ETW brane}
%----------------------------------------

We would like to find an asymptotically AdS$_7\times\mathbb{S}^4$ geometry that ends on an ETW brane in an $SO(5)$-invariant way, and for which the ETW brane is dual to a conformally invariant boundary condition. Such a geometry is invariant under an $SO(2,5)\times SO(5)$ isometry and is constrained to take the form
\begin{equation}
ds^{2}_7 = e^{2 A(\rho)}\left( \frac{-dt^2 + d\vec{x}^2+dz^2}{z^2}\right) + d \rho^{2} \,,
\end{equation}
together with $\phi = \phi(\rho)$. The seven-form flux then reads
\begin{equation}
    F_{t1\hdots 4z \rho} = \frac{3 e^{-\frac{48}{5}\phi + 6 A}}{z^{6}}\,.
\end{equation}
We then proceed to find boundary conditions at $\rho=\bar{\rho}$ such that, with suitable boundary terms, the SUGRA effective action has a consistent variational principle. The calculation goes in the same way as in the discussion around~\eqref{E:delta11bdy}. Imposing a Neumann-like boundary condition on the RR potential at $\rho = \bar{\rho}$ implies the existence of a uniform density of $-N$ units of M5-brane charge smeared homogeneously over the $\mathbb{S}^4$. We posit that this distribution of M5 brane charge is made up by a uniform distribution of $\overline{\text{M}5}$ branes, which leads to a particular boundary contribution to the stress-energy. Including it we then impose a Neumann condition on the metric, which includes the stress tensor of the brane distribution. If we let $\Sigma = \text{vol}_{\mathbb{S}^4}=d^4\theta \,\sqrt{\tilde{g}}$ denote the smearing form, then the total boundary term, the sum of the Gibbons-Hawking term and the action of this distribution of $\overline{\text{M}5}$-branes, reads
\begin{align}
    S_{\rm bdy}& = \frac{1}{2 \kappa_{11}^2} \bigg(2 \int d^{10} x \sqrt{-H}\, K(H)
    \\ 
    \nonumber& \quad -3 \left(\int d^{6} \sigma \, \sqrt{-\text{P}[G]} \wedge \Sigma + \int C_{6} \wedge \Sigma \right) \bigg)\,,
\end{align}
where $\Sigma = \text{vol}_{\mathbb{S}^4} = d^4\theta \sqrt{\tilde{g}}$ is the smearing form with $\tilde{g}$ the metric on a unit $\mathbb{S}^4$.

The Neumann condition on the metric at $\rho=\bar{\rho}$ reads
\begin{align}
\begin{split}
\label{IJCondition}
     K(H)_{\mu \nu} - K(H) H_{\mu \nu}& = -\frac{3}{2} e^{-4\phi} H_{\mu \nu}\,, 
     \\ 
    K(H)_{\alpha \beta} - K(H) H_{\alpha \beta} &= 0\,.
\end{split}
\end{align}
Using the trace of the first line, the angular part implies that the $\overline{\text{M}5}$ branes are in equilibrium. In terms of the quantities $A$ and $\rho$ the junction conditions become
\begin{align}
\begin{split}
\label{E:AdS7israel}
         A' & = - \frac{3}{10}e^{-\frac{24}{5}\phi}\,,
         \\
         \phi' & = -e^{-\frac{24}{5}\phi}\,,
 \end{split}
\end{align}
at $\rho=\bar{\rho}$.

As in our discussion around~\eqref{E:M2interface} we can also arrive at these boundary conditions and the 5-brane part of the ETW brane action by first considering a sharp interface connecting two asymptotically AdS$_7\times\mathbb{S}^4$ regions in a $SO(2,5)\times SO(5)$ invariant way, one supported by $N_+$ units of flux and the other by $N_-$ units, and then taking the $N_-\to 0$ limit.

%----------------------------------------
\subsubsection{Solution}
%----------------------------------------

We seek to find an asymptotically AdS$_7\times\mathbb{S}^4$ solution consistent with the ETW boundary condition~\eqref{E:AdS7israel}. Again it is convenient to solve the $uu$ component of the Einstein's equations in~\eqref{GeneralEOMSAdS7S4}, where $u^{\mu}\partial_{\mu} = z e^{-A}\partial_t + \partial_{\rho}$, along with the scalar equation of motion. These are
\begin{align}
\begin{split}
\label{AdS7EOMs}
    A'' -e^{-2 A} &= -\frac{36}{25} \phi'^{2}\,, 
    \\
     \phi'' + 6 A' \phi' &= 3 e^{-\frac{18}{5} \phi} \left( 1 - e^{-6 \phi} \right)\,.
\end{split}
\end{align}
Using the junction condition~\eqref{E:AdS7israel} the $\rho\rho$-component of Einstein's equations imposes a further constraint on the data at $\rho=\bar{\rho}$,
\begin{equation}\label{ExtraBC}
    e^{-2 A} = \frac{2}{5} e^{-\frac{18}{5} \phi }\,.
\end{equation}
The data $(A,\phi',A')$ at $\rho=\bar{\rho}$ are then determined in terms of $\phi$ there which acts as an initial condition for our numerical evolution.

\begin{figure}[t]
\begin{center}
\includegraphics[width=3in]{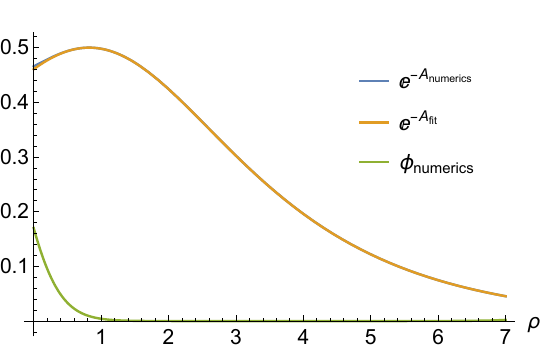} 
\end{center}
\caption{\label{F:AdS7} The numerical profiles for $\phi(\rho)$ (green) and $e^{-A(\rho)}$ (blue) for our asymptotically AdS$_7\times\mathbb{S}^4$ geometry that ends on a distribution of $\overline{\text{M}5}$ branes at $\rho=0$. The 7d geometry is quite nearly empty AdS$_7$ cut off by a tensionful brane of tension $T\approx 0.974$ as indicated by the fit (orange).}
\end{figure}

We then shoot from the ETW brane subject to these boundary conditions and tune the initial condition $\phi(\bar{\rho})$ so as to achieve the asymptotically AdS$_7\times\mathbb{S}^4$ boundary condition $\phi\to 0$ as $\rho\to \infty$. We have found a solution, presented in Fig.~\ref{F:AdS7}, which accomplishes this for fairly large $\rho\lesssim 7$. For that solution we have also computed
\begin{equation}
    \phi_{b} = \int_{0}^{\infty} d \rho\, e^{-A(\rho)} \approx 2.26 < \pi\,,
\end{equation}
which implies that the ETW brane and the AdS$_{7}$ boundary are connected by null geodesics. As with our other examples the noncompact part of the geometry is quite nearly pure AdS, cut off by a tensionful brane of tension $T\approx 0.974$. 

We have also studied the effective potential for a single $\overline{\text{M}5}$ brane separated from the rest of the distribution. As in our examples above its effective potential for $\rho$, provided that it is extended along an AdS$_6$ slice and sitting at a fixed angle on the $\mathbb{S}^4$, is globally maximized at $\rho=\bar{\rho}=0$. As a result this ETW brane is also unstable to fragmentation.

\end{document}